\documentclass[journal=jacsat,manuscript=article]{achemso}
\pdfoutput=1
\usepackage[numbers]{natbib}
\usepackage{chemformula} 
\usepackage[T1]{fontenc} 
\usepackage{amssymb,amsmath,mathtools}
\usepackage[
colorlinks=true,        
allcolors = black,  
]{hyperref}
\hypersetup{
    colorlinks=true,
    linkcolor=black,
    filecolor=black,      
    urlcolor=black,
    pdftitle={Overleaf Example},
    pdfpagemode=FullScreen,
    }

\title[An \textsf{achemso} demo]
  {Highly nonlinear Moir\'e exciton and trion polaritons}	

\author{Arnab Barman Ray}
\affiliation{
 Laboratory for Physical Sciences, University of Maryland, 8050 Greenmead Dr, College Park, MD 20740
}%
\alsoaffiliation{The Institute of Optics, University of Rochester, 480 Intercampus Dr, Rochester, New York 14627, United States}
\author{Trevor Ollis}%
\affiliation{%
Department of Physics and Astronomy, University of Rochester, Rochester, New York 14627, United States
}

\author{Fei Cheng}%
\affiliation{Notre Dame Nanofabrication Facility, University of Notre Dame, Notre Dame, Indiana 46556, United States}

\author{Adam L. Friedman}
\affiliation{
 Laboratory for Physical Sciences, University of Maryland, 8050 Greenmead Dr, College Park, MD 20740
}%

\author{Aubrey T. Hanbicki}
\affiliation{
 Laboratory for Physical Sciences, University of Maryland, 8050 Greenmead Dr, College Park, MD 20740
}%

\affiliation{
 Laboratory for Physical Sciences, University of Maryland, 8050 Greenmead Dr, College Park, MD 20740
}%

\author{Anthony Nickolas Vamivakas}
\affiliation{The Institute of Optics, University of Rochester, 480 Intercampus Dr, Rochester, New York 14627, United States}%
\alsoaffiliation{
 Center for coherence and quantum optics, Department of Physics, University of Rochester, 480 Intercampus Dr, Rochester, New York 14627, United States
}%
\alsoaffiliation{%
Materials Science, University of Rochester, Rochester, New York 14627, United States
}%
\alsoaffiliation{%
Department of Physics and Astronomy, University of Rochester, Rochester, New York 14627, United States
}
\email{nick.vamivakas@rochester.edu, abarmanr@ur.rochester.edu}

\date{\today}
\abbreviations{IR,NMR,UV}
\begin{document}
\begin{abstract}

Moir\'e multi-layers of transition metal dichalcogenides have been shown to exhibit optical responses that are endowed with a richness that is absent in single monolayers. Much of this can be attributed to the Moir\'e superlattice that modulates the electronic landscape of these heterostructures. Strongly coupled layer-hybridized excitons in $\text{MoSe}_2 / \text{WS}_2$ heterobilayers have been shown to exhibit enhanced optical nonlinearities. In this work we strongly couple layer hybridized excitons and trions in n-doped $\text{MoSe}_2 / \text{WS}_2$ heterobilayers inside an optical microcavity. We find that the additional Lindhard screening from dopant electrons and the formation of trions result in a strikingly non-monotonic nonlinear response. The absence of electron capture in the Moir\'e superlattice plays a crucial role, promising very large second-order nonlinearities. In this work, trion polaritons manifest as high velocity hot polaritons, reaching nominal diffusion lengths approaching 100 microns.
\end{abstract}

\maketitle

\section{Introduction}
Exciton-polaritons have numerous potential applications, ranging from energy harvesting \cite{1,2} and unconventional lasing \cite{3,4,5} to quantum computing gates \cite{6,7} and all-optical switching \cite{8,9,46,47}, as well as being testbeds for quantum many body physics \cite{10}. In recent years, Moir\'e superlattices made of monolayers of transition metal dichalcogenides have revealed a variety of rich physical phenomenon \cite{11,12,13,14,15,16}. Research has progressed to multilayer Moir\'e heterostructures that host intriguing physics \cite{17,18}, such as relativistic Mott insulators and quadrupolar excitons. For the simplest case of a bilayer, when two monolayers are brought in contact with a small angle of twist or with lattice mismatches, a superlattice forms on top of the crystal lattices of the individual monolayers \cite{19}. This superlattice serves as a periodic potential that can introduce flat bands and spatially localize bound states such as excitons and trions \cite{20}. In some cases, these trapped states have been shown to emit single photons \cite{21}, exhibit large electrical tunability, and host other unconventional behaviors such as many-exciton interaction-induced Zeeman effects \cite{43} that are absent in the monolayers. One particularly interesting feature that has been discovered for interlayer excitons in a Moir\'e superlattice is the complete absence of electron capture which results in a strong phase-space filling effect for trions \cite{22}. This raises the possibility of leveraging these nonlinear features of trions in a strongly coupled system to induce nonlinear photon-photon interactions \cite{23,24}. Recently, Moir\'e excitons in bilayers of $\text{MoSe}_2/\text{WS}_2$ were strongly coupled in a microcavity and have been shown to exhibit strong nonlinearities owing to the 0D nature of the excitons and exciton blockade effects pertaining to a maximum of one exciton occupation at each Moir\'e site \cite{25}. In the same system, electrical gating has provided evidence for a Mott insulating polariton state at an integer filling of $\nu=1$ \cite{26} (one electron per Moir\'e site). 

In this work, we investigate an n-doped bilayer of $\text{MoSe}_2/\text{WS}_2$ at a twist angle of $58.35^\circ\pm2.14^\circ$. Inside an optical microcavity, we investigate the hyperspectrum of the cavity-coupled excitonic states at cryogenic temperatures of 5K. We find evidence for strongly coupled Moir\'e trions. By investigating the polariton density dependent characteristics of the hyperspectrum, we find a nonlinear, non-monotonic dependence of the oscillator strengths of the species involved. Using a phenomonological model for Lindhard screening by charge carriers \cite{28}, we find that the observations can be accurately described by considering the initial depletion of dopant carriers (and hence the screening density) through trion formation and the absence of trions formed from optically excited carriers through electron capture.  Throughout the polariton density range investigated, the lower polariton branch is seen to exhibit an energy shift of $\sim 4\,\text{meV}$. Temperature-dependent measurements confirm the trion origin of the upper polariton branch as it disappears at temperatures around $30\,\text{K}$. Finally, looking at real space spectrum of the polariton emission, we show that the strongly coupled trions near a relaxation bottleneck propagate very efficiently, reaching nominal values of diffusion lengths close to $100\,\mu \text{m}$.

\section{Methods}
\subsection{Sample fabrication}
The n-doped monolayers and hBN (hexagonal boron nitride) were mechanically exfoliated from high-quality bulk samples obtained from 2D Semiconductors. The doping densities for the TMDC crystals are reported to be $10^{18}\,\text{cm}^{-3}$. The individual flakes were then assembled step-by-step under an optical microscope using dome-shaped windows constructed from cured PDMS (Poly-dimethylsiloxane) with a thin pane of PPC (poly-propylene carbonate). After the device stacking, the assembly was heated to release it when in contact with a distributed Bragg reflector (DBR) chip (with 98 nm of $\text{SiO}_2$ on top). The SiN-terminated DBR was fabricated using a PECVD (plasma-enhanced chemical vapor deposition) method, with 10.5 pairs of SiN/$\text{SiO}_2$. After the deposition of the stack, the sample was spin-coated with PMMA and finally a silver mirror of thickness $\sim 55$ nm was deposited on top through e-beam deposition. The thicknesses of the hBN, the $\text{SiO}_2$ layer and the PMMA were optimized to spectrally position the cavity resonance near the excitons in the sample.
\subsection{Optical measurements}
The measurements were carried out using a custom-built confocal microscope. A Coherent Chameleon titanium-sapphire laser was used in its mode-locked configuration at $725$ nm for the photoluminescence (PL) experiments on the microcavity. The wavelength used for the PL data on the exposed bilayer was $730$ nm. The short pulse width (~1 ps) and the repetition rate (80 MHz) ensured effective heat dissipation even at larger fluences. The laser was focused to a submicrometer-diameter spot using a 0.70 NA objective lens in a closed-cycle cryostat (Montana Instruments) at $\sim 5\, \text{K}$. The emission spot is relayed through the objective and a Fourier lens (air-spaced achromat) and imaged onto the CCD camera using an achromatic lens system. The collected PL spectrum is then analyzed using a Princeton Instruments spectrometer (Acton SP-2750i) and an $\text{LN}_2$ cooled Pylon CCD camera. Reflectance measurements were carried out using a fiber coupled Thorlabs MCWHF2 and MC780 white light sources. Reflectance measurements over longer wavelength ranges were obtained using a Filmmetrics thin film analyzer using a silicon reference. For the real space measurement the Fourier lens was replaced by a collimating lens that sent the emission spot image to the relay system. The Chameleon laser was used at $800$ nm for the second-harmonic generation measurements. 

\begin{figure}
   \includegraphics[scale=0.8]{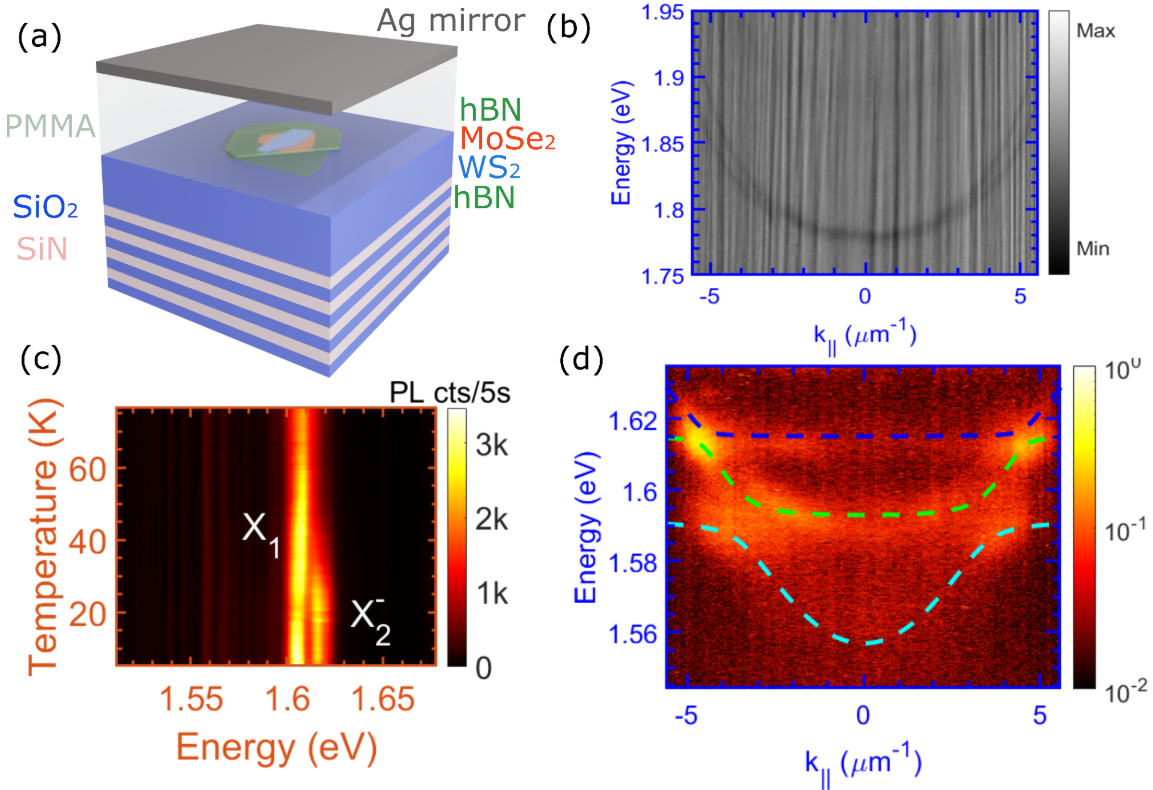}
   \caption{(a) Render of the device geometry (thickness are not representative) with the bottom DBR, $\text{SiO}_2$ and PMMA spacer layers and a top silver mirror, (b) cavity dip away from the heterobilayer, (c) temperature dependent PL spectra from the bare sample without the top mirror (532 nm CW excitation) exhibiting the exciton $X_1$ and the trion $X_2^-$, (d) normalized Fourier spectra of the heterobilayer PL inside the cavity exhibiting the upper polariton (indigo), middle polariton (green) and lower polariton (cyan), 725 nm pulsed (1 ps) ti-sapphire excitation at an average polariton density of $77\,\mu m^{-2}$.}
   \label{fig1}
\end{figure}

\section{Results}
The device is exhibited in Fig.~\ref {fig1}(a). The bare cavity reflectace exhibits a sharp dip at an energy of $1.781 \,\text{eV}$, with a linewidth, $\gamma_c = 4.591\pm0.48 \,\text{meV}$. On top of the two hBN, this dip redshifts to $1.570\,\text{eV}$, due to the increased optical path and higher refractive index of the hBN. The bare sample, without the top mirror showcases the typical resonances  [see Supporting Fig. 4(b)] associated with the layer hybridized excitons $X_1$ and $X_2$, as identified elsewhere \cite{27}. These excitons are composed of a hole in the topmost valence band of $\text{MoSe}_2$, and an electron hybridized across the conduction bands in the two monolayers. In addition, we also see signatures of the $X_2$ trion, $X_2^-$, which is identified from its temperature-dependent PL decay in Fig.~\ref{fig1}(c). Once the cavity is formed, the Fourier image of the PL clearly indicates the strong coupling of two of the three resonances, $X_1$ and $X_2^-$ to the cavity photon. The topmost polariton branch, which arises from coupling with the $X_2$ exciton, is not visible in the PL spectra, possibly due to fast relaxation processes \cite{29}. The observed dispersions can be explained with a three-resonance model involving the two excitons and the single trion. We diagonalize the coupled oscillator model Hamiltonian, 
\begin{equation}
    H = \begin{bmatrix}
    E_{X_2} & 0 & 0 & \Omega_{X_2} \\
    0 & E_{X_2^-} & 0 & \Omega_{X_2^-} \\
    0 & 0 & E_{X_1} & \Omega_{X_1} \\
    \Omega_{X_2} & \Omega_{X_2^-} &  \Omega_{X_1} & E_c 
    \end{bmatrix}
\end{equation}
with the resonant energies and the cavity resonance, and their respective coupling strengths. At an average polariton density of $\sim 77\, \mu m^{-2}$ (see supporting information section III for details), we extract the energies,  $E_{X_2} = 1.649 \,\text{eV}$, $E_{X_2^-} = 1.615 \,\text{eV}$, $E_{X_1} = 1.592 \,\text{eV}$  and $E_c = 1.570 \,\text{eV}$, and the coupling strengths (Rabi splittings), $\Omega_{X_2} = 32.42\pm 0.25 \,\text{meV}$, $\Omega_{X_2^-} = 3.27\pm 0.25 \,\text{meV}$ and $\Omega_{X_1} = 7.46\pm 0.20 \,\text{meV}$. The extracted energies are within a few meV of what is observed in the uncoupled heterobilayer. We note that we observed a rather large Rabi splitting for the exciton $X_2$, previously reported values for which were closer to about $10 \,\text{meV}$. This arises from the dependence of the oscillator strengths of the two excitonic species on the twist angle. In fact, it has been observed that as the angle of twist changes from $0^\circ$ to $60^\circ$, the oscillator strength is redistributed from $X_1$ to $X_2$ \cite{27}. The extracted coupling strength for the trion is small; however, the trion linewidth extracted from PL data which is $\gamma_{X_2^-} = 4.92 \pm 0.30\,\text{meV}$ puts the interaction squarely in the strong coupling regime, since $\Omega_{X_2^-} > (\gamma_{X_2^-} + \gamma_{c})/4$  \cite{30}.

\begin{figure}
   \includegraphics[scale=0.50]{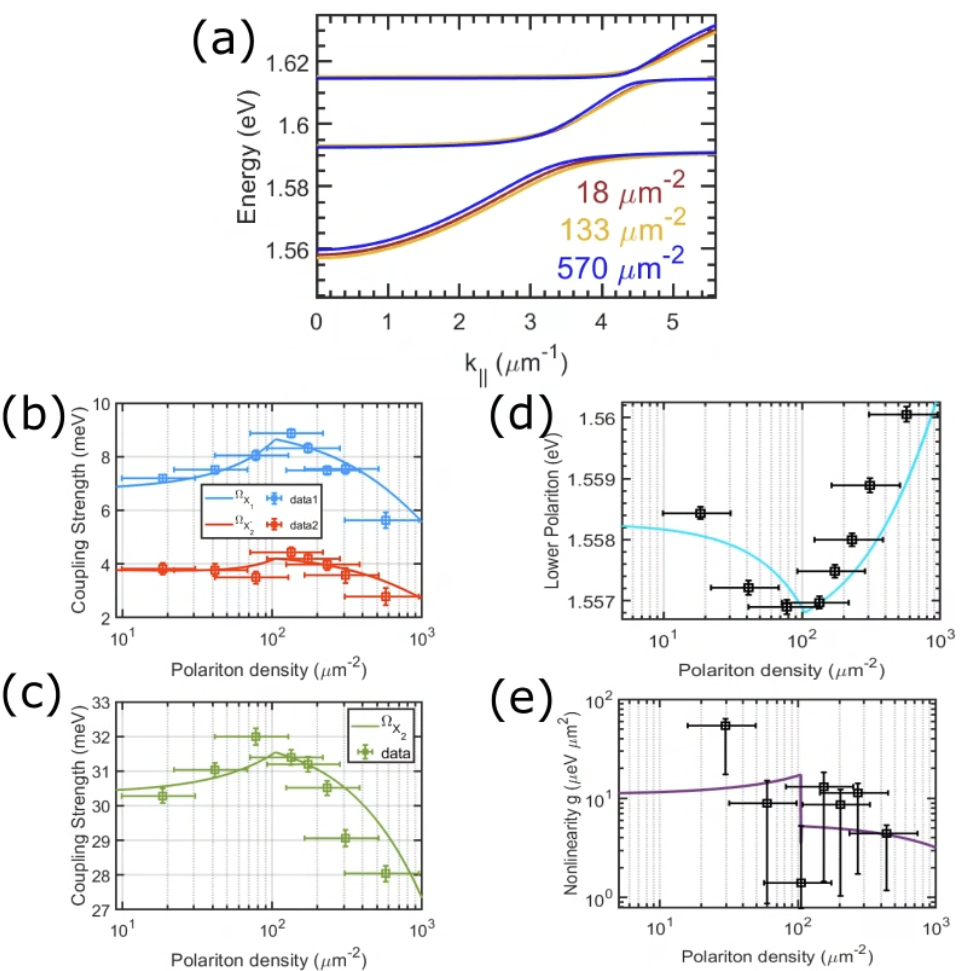}
  \caption{(a) Fitted PL dispersions of the three polariton branches, color coded according to their polariton densities. Coupling strengths of the (b) $X_1$ exciton and $X_2^-$ trion, and (c) $X_2$ exciton as a function of polariton density. The solid lines are fits obtained from the Lindhard screening model. (d) Variation in the lower polariton energy at $k_{||} = 0$ as a function of polariton density, with the fit. (e) Theoretical nonlinearity curve and the extracted nonlinearities (displayed at the average polariton densities, between the two data points used). Errors in coupling strengths and energies are standard errors.}
  \label{fig2}
\end{figure}

We study the evolution of the polariton dispersion with polariton density in in Fig. ~\ref{fig2}(a). Our pumping scheme is a near-resonant one, with the exciting laser (725 nm/1.71 eV) lying above the topmost polariton branch ($\sim 1.649\,\text{meV}$) \cite{39,40}. In our analysis, we fix the cavity resonance energy, and we allow the resonance energies to only vary over a very small window of $1\,\text{meV}$, as 0-dimensional excitons are known to exhibit constant dephasing with changing polariton density \cite{25}. From these constraints, we extract the variation of the polariton energies and oscillator/coupling strengths as a function of polariton density.  We find that over the density range investigated, the lower polariton branch undergoes a non-monotonic energy shift, initially decreasing in energy to reach a minimum near $100\, \mu m^{-2}$. We then observe a dramatic change from this redshift to the previously observed\ blueshift with increasing density. Overall, we observe a nonlinear change in the energy of this polariton branch by about $3.3\,\text{meV}$, larger than previous reports \cite{23,24,25} in similar systems. The extracted coupling strengths for all three species involved also exhibit this unusual non-monotonic behavior as exhibited in Fig.~\ref{fig2}(b) and (c), increasing up to a certain polariton density and decreasing thereafter. We note that this non-monotonic, nonlinear behavior has never been reported previously, which led us to conclude that the Moiré superlattice must be playing a crucial role.

To understand this behavior, we looked at the factors that can affect oscillator strengths of resonances in semiconductor systems. One of the physical determinants of the oscillator strength is the screening experienced by the exciton or trion in question from the background of charge carriers and other excitons/trions through Coulomb blockade. However, these effects are generally monotonic, with oscillator strengths decreasing with increasing polariton/carrier density. In order to account for this unusual observation, we must turn to the absence of electron capture observed in some Moiré systems \cite{22}. Since we start out with monolayers that are n-doped, there is an initial density of doped electron carriers that must serve to screen the electromagnetic interactions of excitons and trions. However, with increasing optical excitation density, the density of these dopant electrons must reduce due to trion formation. This effect can then lead to an initial increase in the oscillator strengths up to a certain polariton density that is dependent on the local dopant density of the location being investigated. In fact, in our sample, we do find that the oscillator strengths vary across the sample region. To quantify this effect, we turn to a rate equation and Lindhard screening-based phenomenological model. In contrast to intralayer excitons in monolayers, these layer hybridized excitons and trions are expected to possess non-zero permanent electrical dipole moments, making them more sensitive to such screening. Taken together with excitonic Coulomb blockade \cite{25}, we find a suitable explanation of the observed behavior.

We start with the rate equations, 
\begin{align}
    \frac{dN_{X_1}}{dt} = \eta G - \frac{N_{X_1}}{\tau_{X_1}},\\
    \frac{dN_{X_2}}{dt} = \eta G - \frac{N_{X_2}}{\tau_{X_2}}-k_eN_{X_2}n_e-k_pN_{X_2}n_p,\\
    \frac{dN_{X_2^-}}{dt} = k_eN_{X_2}n_e + k_eN_{X_2}n_p -\frac{N_{X_2^-}}{\tau_{X_2^-}},
\end{align}
where $\eta$ is the generation rate for excitons, $G$ is the excitation density, $N_X$ and $\tau_X$ are the density and lifetime of species 'X' respectively, '$k_e$' is the exciton-to-trion conversion rate, and $n_e = n_e^0-N_{X_2^-}$ is the remaining dopant density, where $n_e^0$ is the dopant density at zero excitation. $n_p$ is the density of photogenerated free carriers, which we take to be proportional to excitation through a constant $\eta^\prime, n_p = \eta^\prime G$, with $k_p$ being the trion formation rate through free carriers. Despite the excitation energy ($1.71$ eV) being below the bandgap, a small density of free carriers at elevated excitonic densities is to be expected due to two-photon absorption and excitonic dissociation. The fitting process renders the rate of trion formation through free carriers negligible compared to $k_e$, that is, $k_e >> k_p$, highlighting the absence of optically generated trions through electron capture in Moir\'e superlattices \cite{22}. The above equations can be solved for the quasi-steady state conditions of our experiments (since we use a pulsed laser) to give involved expressions for the densities of $X_1$, $X_2$, and $X_2^-$. Expressions for the persisting dopant density $n_e$ can also be obtained.

Lindhard screening, the theory of dynamic screening by an electron gas gives the two well-known limits of Thomas-Fermi (static) screening and the Drude model at the static and long-wavelength limits, respectively. However, neither of these is adequate to describe screening of excitons in TMDCs due to the large binding energies involved \cite{28,31,32,33}. For a two-dimensional gas of electrons, the effect of carrier screening on the exciton radius\cite{41,42} is included in our model by using a phenomenological term $a_B \propto (1+\beta n_s)$, where $a_B$ is the Bohr radius of the exciton, $n_s$ is the carrier density involved in screening, given as $n_s = max(0,n_e^0-N_{X_2^-})+n_p$, and $\beta$ is a screening length. The oscillator strength of an excitonic species is dependent on the overlap of the electron and hole wavefunctions, and hence inversely on the square of the excitonic radius. Hence, we include the effect, $\Omega_X\propto \frac{1}{(1+\beta_X n_s)^2}$. This is added to the familiar excitonic Coulomb blockade effect, $\Omega_X \propto  \frac{1}{\sqrt{(1+\frac{n_X}{n_S})}}$ (where $n_X$ is the total exciton density and $n_S$ is the saturation density of excitons) to arrive at:
\begin{equation}
\Omega_X(n) =  \frac{\Omega_X^0}{\sqrt{(1+\frac{n_X}{n_S})}(1+\beta_X n_s)^2}
\label{eq3}
\end{equation}

As the excitation density increases, there is a gradual decrease in the density of dopant electrons from trion formation, which reaches a minimum at a specific polariton density that is dependent on the local dopant density of electrons. Hence, even though Eq.~(\ref{eq3}) is monotonic in carrier density ($n_s$) and exciton density ($n_X$), it still gives rise to a non-monotonic change in the observed oscillator strengths. Figure~\ref{fig2}(b) and (c) display the theoretical fits to the observed data for the two excitons and the trions. From the fits, we obtain the screening coefficients, $\beta_{X_1} = (2.50\pm0.44)\times10^{-12}\,\text{cm}^{-2}$, $\beta_{X_2} = (0.46\pm0.10)\times10^{-12}\,\text{cm}^{-2}$, and $\beta_{X_2^-} = (4.18\pm1.05)\times10^{-12}\,\text{cm}^{-2}$. The larger screening coefficient for the trion aligns well with our expectation that a charged complex would experience larger net screening from free carriers. The fit value of the exciton saturation density is $n_S = (3.81\pm1.45) \times 10^{12}\, \text{cm}^{-2}$.

The energy shift of the lower polariton branch is displayed in Fig.~\ref{fig2}(d), alongside the theoretical curve obtained from the model. For the lower polariton, we calculate a nonlinear coefficient, $g(n) = \frac{dE_{LP}}{dn} = 53.73+(-)\,9.40\,(36.39) \, \mu eV\mu m^2$ (calculated at an average polariton density of $18.49\, \mu m^{-2}$), which is comparable to nonlinearities reported for interlayer excitons in bilayer $\text{MoSe}_2$ \cite{23}, and are some of the highest reported on a TMDC-based polaritonic platform. The calculated $g(n)$ is $\sim7$ times larger than Moir\'e exciton-polaritons in undoped heterobilayers which have exhibited $g(n) \sim 6\, \mu eV\mu m^2$ at similar densities \cite{25}. Trion-polaritons in monolayers have been shown to exhibit $g(n) \sim 30\, \mu eV\mu m^2$ \cite{24}. Errors in polariton densities are defined by upper and lower bounds, which are calculated by accounting for standard/instrumental errors in the parameters/data used (see supporting information section III). The nonlinear coefficients calculated and the expected theoretical curve are displayed in Fig.~\ref{fig2}(e). For larger polariton densities, we report nonlinear coefficients similar (within the same order of magnitude) to those obtained for undoped heterobilayers in a previous work \cite{25}. This arises from the fact that the heterobilayer is essentially undoped under intense excitation. However, at lower polariton densities, the doped heterobilayer exhibits a jump in the nonlinear coefficient yielding nonlinearities that are several times larger than what has been observed for undoped heterobilayers. Strikingly, we see that the model used predicts a theoretically `infinite' second-order nonlinearity at a certain polariton density that depends on the local dopant density.

\begin{figure}[h]
   \includegraphics[scale=0.5]{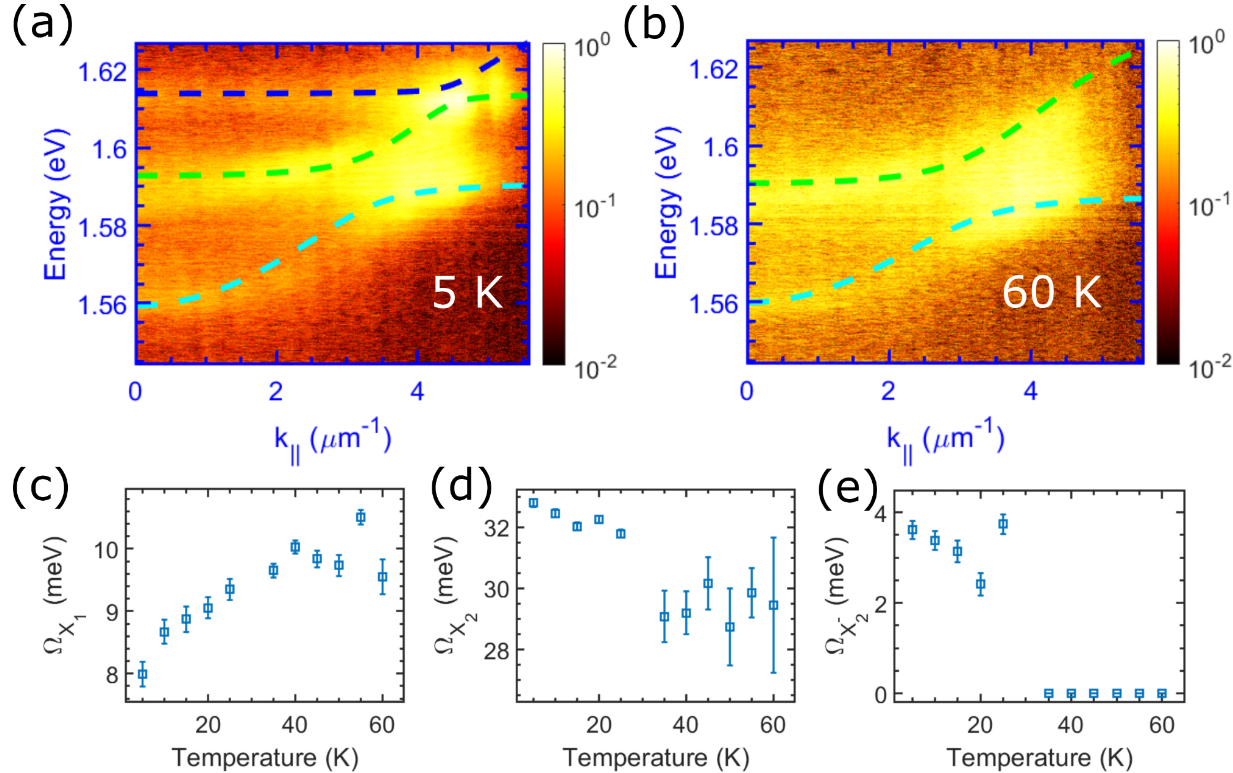}
   \caption{(a) PL dispersion at a temperature of 5K, (b) dispersion at 60 K. Temperature dependence of coupling strengths of the (c) $X_1$ exciton, (d) $X_2$ exciton and (e) $X_2^-$ trion. The polariton density in all cases is at $\sim 1.36\times10^3\, \mu m^{-2}$. }
   \label{fig3}
\end{figure}

The experienced nonlinearity and the critical polariton density (where the sharp reversal occurs) have been observed to vary across the sample, resulting from a nonuniform dopant density. At sites where the electron density is higher, we see that trions remain strongly coupled even at elevated polariton densities. We investigate the temperature dependence of the polariton dispersions in Fig.~\ref{fig3}, where the data was collected from a different spot in the sample. As expected, we see that the trion branch disappears completely above $30\, \text{K}$ in Fig.~\ref{fig3}(b). We note that $\Omega_{X_1}$ generally increases with temperature up to 40 K in Fig.~\ref{fig3}(c), an effect that which has been observed before \cite{25}. This initial increase is driven by an increase in the bare exciton linewidth, with a constant inhomogeneous term, along with a temperature-dependent homogeneous broadening term that would include effects of scattering through acoustic and LO phonons \cite{38}. In Fig.~\ref{fig3}(d) and (e), $\Omega_{X_2}$ and $\Omega_{X_2^-}$ are seen to generally decrease with increasing temperature, suggesting a substantial difference in phonon interaction across the two species of excitons investigated

\begin{figure}
   \includegraphics[scale=0.60]{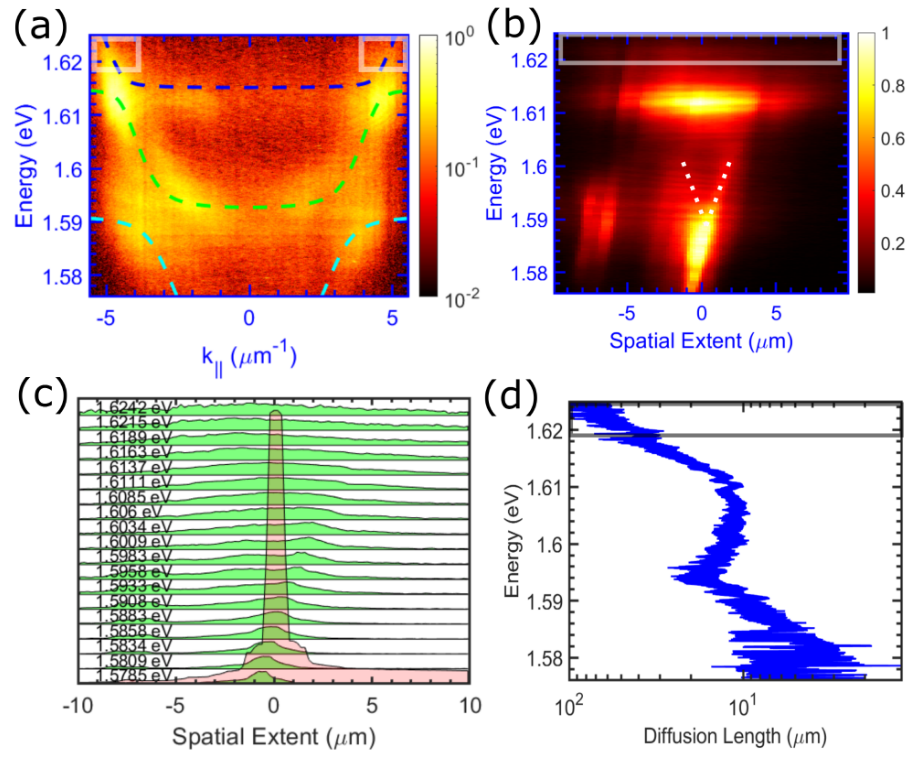}
    \caption{(a) PL dispersion at a temperature of 5K and polariton density of $\sim14.79\,\mu m^{-2}$. (b) Spatio-spectrum of the PL emission, with travelling middle polariton branches indicated with white dotted lines. (c) Peak-normalized spectral slices of the real space image at different energies exhibiting extended diffusion at energies corresponding to the bottleneck trion-polariton branch (spatial profile of the 725 nm pulsed laser is indicated by the red shaded area), (d) extracted nominal diffusion lengths as a function of energy. The high-velocity trion-polaritons are indicated in (a), (b), and (d). }
    \label{fig4}
\end{figure}

Finally, to firmly establish the strongly coupled nature of the trion-polaritons, we examine the real-space spectrum of the PL emission. At a low polariton density of $\sim 8\,\mu m^{-2}$, we display the PL dispersion in Fig.~\ref{fig4}(a), where the large velocity trion polaritons are outlined in the white boxes. If we image the emission spot on the spectrometer, we acquire the spatio-spectrum as shown in Fig.~\ref{fig4}(b). The data exhibits the spatial movement of the travelling middle polaritons, as denoted by the white dotted straight lines. Linecuts of the spatio-spectrum at different energies are presented in Fig.~\ref{fig4}(c). Here, from the peak-normalized slices, we can see that the spatial tails extend further as we approach the energies of the trion polaritons, thus confirming their large group velocities. To quantify this diffusion, we define a nominal diffusion length through the equation $n(r) \propto \int_{-\infty}^{\infty} dx^\prime K_0 \left(\frac{x^\prime}{L_D}\right)e^{-\frac{(r-r^\prime)^2}{L^2}}$, where $K_0$ is the modified Bessel function of the second kind \cite{34,35,36,37}, $r$ is the spatial coordinate, $L$ quantifies the spatial extent of the gaussian laser spot, and $L_D$ is the obtained diffusion length, which is displayed in Fig.~\ref{fig4}(d). We see that the trion polaritons in our sample reach nominal diffusion lengths close to $10^2\,\mu m$, which would be impossible for uncoupled localized trions\cite{44,45}.

\section{Conclusion}
To summarize, we strongly couple resonances in a Moir\'e heterobilayer inside a metal-DBR microcavity. We demonstrate strongly coupled Moir\'e trions. Since we use n-doped monolayers, it gives rise to a unique non-monotonic dependence of the oscillator strengths on polariton density, which we ascribe to a Moir\'e superlattice-induced effect that leads to the preferential formation of trions from dopant electrons. The layer hybridized nature of the excitons can also contribute to this observed phenomenon through their non-zero permanent electric dipole moments.

We quantify polariton nonlinearities that are larger than any similar systems that have been explored previously, at $ 53.73+(-)\,9.40\,(36.39) \, \mu eV\mu m^2$. Our model predicts an unusually large second-order polariton nonlinearity that may be controlled through electrical doping in future devices, allowing for unprecedented control in photon-photon nonlinearity, perhaps all the way down to a single photon/dopant electron \cite{46,47} - densities that are still an order of magnitude lower. Our results pave the way for polariton-based devices that exploit nonlinear interactions.

\section{Acknowledgments}
This work was supported by AFOSR FA9550-19-1-0074 from the Cornell Center for Materials Research. The authors thank URNano for the use of their facilities. This project was supported in part by an appointment to the NRC Research Associateship Program at the Laboratory for Physical Sciences, administered by the Fellowships Office of the National Academies of Sciences, Engineering, and Medicine.

\section{Author Contributions}
A.N.V. conceived and guided the project. F.C. and A.B.R. fabricated the devices. A.B.R. led the experiments with the help of T.O., A.T.H. and A.L.F.. A.B.R. curated and analyzed the data. All authors contributed equally to the discussion and the final manuscript.

\appendix

\bibliography{apssamp}

@PREAMBLE{
 "\providecommand{\noopsort}[1]{}" 
 # "\providecommand{\singleletter}[1]{#1}%" 
}

@article{1,
   author = {Silori, Yogita and Liu, Bin and Li, Yongxi and Forrest, Stephen R. and Ogilvie, Jennifer P.},
   title = {Impact of cavity strong coupling on the charge transfer dynamics in organic donor-acceptor heterojunctions},
   journal = {Physical Review B},
   volume = {111},
   number = {23},
   pages = {235119},
   DOI = {10.1103/859s-sc6n},
   url = {https://link.aps.org/doi/10.1103/859s-sc6n},
   year = {2025},
   type = {Journal Article}
}

@article{2,
   author = {Wang, Mao and Hertzog, Manuel and Börjesson, Karl},
   title = {Polariton-assisted excitation energy channeling in organic heterojunctions},
   journal = {Nature Communications},
   volume = {12},
   number = {1},
   pages = {1874},
   ISSN = {2041-1723},
   DOI = {10.1038/s41467-021-22183-3},
   url = {https://doi.org/10.1038/s41467-021-22183-3},
   year = {2021},
   type = {Journal Article}
}

@article{3,
   author = {Zhang, Long and Hu, Jiaqi and Wu, Jinqi and Su, Rui and Chen, Zhanghai and Xiong, Qihua and Deng, Hui},
   title = {Recent developments on polariton lasers},
   journal = {Progress in Quantum Electronics},
   volume = {83},
   pages = {100399},
   ISSN = {0079-6727},
   DOI = {https://doi.org/10.1016/j.pquantelec.2022.100399},
   url = {https://www.sciencedirect.com/science/article/pii/S0079672722000258},
   year = {2022},
   type = {Journal Article}
}

@article{4,
   author = {Zhao, Jiaxin and Su, Rui and Fieramosca, Antonio and Zhao, Weijie and Du, Wei and Liu, Xue and Diederichs, Carole and Sanvitto, Daniele and Liew, Timothy C. H. and Xiong, Qihua},
   title = {Ultralow Threshold Polariton Condensate in a Monolayer Semiconductor Microcavity at Room Temperature},
   journal = {Nano Letters},
   volume = {21},
   number = {7},
   pages = {3331-3339},
   ISSN = {1530-6984},
   DOI = {10.1021/acs.nanolett.1c01162},
   url = {https://doi.org/10.1021/acs.nanolett.1c01162},
   year = {2021},
   type = {Journal Article}
}

@article{5,
   author = {Fan, Yuening and Wan, Qiaochu and Yao, Qi and Chen, Xingzhou and Guan, Yuanjun and Alnatah, Hassan and Vaz, Daniel and Beaumariage, Jonathan and Watanabe, Kenji and Taniguchi, Takashi and Wu, Jian and Sun, Zheng and Snoke, David},
   title = {High Efficiency of Exciton-Polariton Lasing in a 2D Multilayer Structure},
   journal = {ACS Photonics},
   volume = {11},
   number = {7},
   pages = {2722-2728},
   DOI = {10.1021/acsphotonics.4c00549},
   url = {https://doi.org/10.1021/acsphotonics.4c00549},
   year = {2024},
   type = {Journal Article}
}

@article{6,
   author = {Kyriienko, O. and Liew, T. C. H.},
   title = {Exciton-polariton quantum gates based on continuous variables},
   journal = {Physical Review B},
   volume = {93},
   number = {3},
   pages = {035301},
   DOI = {10.1103/PhysRevB.93.035301},
   url = {https://link.aps.org/doi/10.1103/PhysRevB.93.035301},
   year = {2016},
   type = {Journal Article}
}

@article{7,
   author = {Ghosh, Sanjib and Liew, Timothy C. H.},
   title = {Quantum computing with exciton-polariton condensates},
   journal = {npj Quantum Information},
   volume = {6},
   number = {1},
   pages = {16},
   ISSN = {2056-6387},
   DOI = {10.1038/s41534-020-0244-x},
   url = {https://doi.org/10.1038/s41534-020-0244-x},
   year = {2020},
   type = {Journal Article}
}

@article{8,
   author = {Ballarini, D. and De Giorgi, M. and Cancellieri, E. and Houdré, R. and Giacobino, E. and Cingolani, R. and Bramati, A. and Gigli, G. and Sanvitto, D.},
   title = {All-optical polariton transistor},
   journal = {Nature Communications},
   volume = {4},
   number = {1},
   pages = {1778},
   ISSN = {2041-1723},
   DOI = {10.1038/ncomms2734},
   url = {https://doi.org/10.1038/ncomms2734},
   year = {2013},
   type = {Journal Article}
}

@article{9,
   author = {Feng, Jiangang and Wang, Jun and Fieramosca, Antonio and Bao, Ruiqi and Zhao, Jiaxin and Su, Rui and Peng, Yutian and Liew, Timothy C. H. and Sanvitto, Daniele and Xiong, Qihua},
   title = {All-optical switching based on interacting exciton polaritons in self-assembled perovskite microwires},
   journal = {Science Advances},
   volume = {7},
   number = {46},
   pages = {eabj6627},
   DOI = {doi:10.1126/sciadv.abj6627},
   url = {https://www.science.org/doi/abs/10.1126/sciadv.abj6627},
   year = {2021},
   type = {Journal Article}
}

@article{10,
   author = {Deng, Hui and Haug, Hartmut and Yamamoto, Yoshihisa},
   title = {Exciton-polariton Bose-Einstein condensation},
   journal = {Reviews of Modern Physics},
   volume = {82},
   number = {2},
   pages = {1489-1537},
   DOI = {10.1103/RevModPhys.82.1489},
   url = {https://link.aps.org/doi/10.1103/RevModPhys.82.1489},
   year = {2010},
   type = {Journal Article}
}

@article{11,
   author = {Chen, Dongxue and Lian, Zhen and Huang, Xiong and Su, Ying and Rashetnia, Mina and Ma, Lei and Yan, Li and Blei, Mark and Xiang, Li and Taniguchi, Takashi and et. al.},
   title = {Excitonic insulator in a heterojunction moiré superlattice},
   journal = {Nature Physics},
   volume = {18},
   number = {10},
   pages = {1171-1176},
   ISSN = {1745-2481},
   DOI = {10.1038/s41567-022-01703-y},
   url = {https://doi.org/10.1038/s41567-022-01703-y},
   year = {2022},
   type = {Journal Article}
}

@article{12,
   author = {Tang, Yanhao and Li, Lizhong and Li, Tingxin and Xu, Yang and Liu, Song and Barmak, Katayun and Watanabe, Kenji and Taniguchi, Takashi and MacDonald, Allan H. and Shan, Jie and Mak, Kin Fai},
   title = {Simulation of Hubbard model physics in WSe2/WS2 moiré superlattices},
   journal = {Nature},
   volume = {579},
   number = {7799},
   pages = {353-358},
   ISSN = {1476-4687},
   DOI = {10.1038/s41586-020-2085-3},
   url = {https://doi.org/10.1038/s41586-020-2085-3},
   year = {2020},
   type = {Journal Article}
}

@article{13,
   author = {Miao, Shengnan and Wang, Tianmeng and Huang, Xiong and Chen, Dongxue and Lian, Zhen and Wang, Chong and Blei, Mark and Taniguchi, Takashi and Watanabe, Kenji and Tongay, Sefaattin and Wang, Zenghui and Xiao, Di and Cui, Yong-Tao and Shi, Su-Fei},
   title = {Strong interaction between interlayer excitons and correlated electrons in WSe2/WS2 moiré superlattice},
   journal = {Nature Communications},
   volume = {12},
   number = {1},
   pages = {3608},
   ISSN = {2041-1723},
   DOI = {10.1038/s41467-021-23732-6},
   url = {https://doi.org/10.1038/s41467-021-23732-6},
   year = {2021},
   type = {Journal Article}
}

@article{14,
   author = {Gu, Jie and Ma, Liguo and Liu, Song and Watanabe, Kenji and Taniguchi, Takashi and Hone, James C. and Shan, Jie and Mak, Kin Fai},
   title = {Dipolar excitonic insulator in a moiré lattice},
   journal = {Nature Physics},
   volume = {18},
   number = {4},
   pages = {395-400},
   ISSN = {1745-2481},
   DOI = {10.1038/s41567-022-01532-z},
   url = {https://doi.org/10.1038/s41567-022-01532-z},
   year = {2022},
   type = {Journal Article}
}

@article{15,
   author = {Zhang, Zuocheng and Regan, Emma C. and Wang, Danqing and Zhao, Wenyu and Wang, Shaoxin and Sayyad, Mohammed and Yumigeta, Kentaro and Watanabe, Kenji and Taniguchi, Takashi and Tongay, Sefaattin and Crommie, Michael and Zettl, Alex and Zaletel, Michael P. and Wang, Feng},
   title = {Correlated interlayer exciton insulator in heterostructures of monolayer WSe2 and moiré WS2/WSe2},
   journal = {Nature Physics},
   volume = {18},
   number = {10},
   pages = {1214-1220},
   ISSN = {1745-2481},
   DOI = {10.1038/s41567-022-01702-z},
   url = {https://doi.org/10.1038/s41567-022-01702-z},
   year = {2022},
   type = {Journal Article}
}

@article{16,
   author = {Xu, Yang and Liu, Song and Rhodes, Daniel A. and Watanabe, Kenji and Taniguchi, Takashi and Hone, James and Elser, Veit and Mak, Kin Fai and Shan, Jie},
   title = {Correlated insulating states at fractional fillings of moiré superlattices},
   journal = {Nature},
   volume = {587},
   number = {7833},
   pages = {214-218},
   ISSN = {1476-4687},
   DOI = {10.1038/s41586-020-2868-6},
   url = {https://doi.org/10.1038/s41586-020-2868-6},
   year = {2020},
   type = {Journal Article}
}

@article{17,
   author = {Ma, Liguo and Chaturvedi, Raghav and Nguyen, Phuong X. and Watanabe, Kenji and Taniguchi, Takashi and Mak, Kin Fai and Shan, Jie},
   title = {Relativistic Mott transition in twisted WSe2 tetralayers},
   journal = {Nature Materials},
   ISSN = {1476-4660},
   DOI = {10.1038/s41563-025-02359-8},
   url = {https://doi.org/10.1038/s41563-025-02359-8},
   year = {2025},
   type = {Journal Article}
}

@article{18,
   author = {Meng, Yuze and Ma, Lei and Yan, Li and Khalifa, Ahmed and Chen, Dongxue and Zhang, Shuai and Banerjee, Rounak and Taniguchi, Takashi and Watanabe, Kenji and Tongay, Seth Ariel and Hunt, Benjamin and Lin, Shi-Zeng and Yao, Wang and Cui, Yong-Tao and Chatterjee, Shubhayu and Shi, Su-Fei},
   title = {Strong-interaction-driven quadrupolar-to-dipolar exciton transitions in a trilayer moiré superlattice},
   journal = {Nature Photonics},
   ISSN = {1749-4893},
   DOI = {10.1038/s41566-025-01741-x},
   url = {https://doi.org/10.1038/s41566-025-01741-x},
   year = {2025},
   type = {Journal Article}
}

@article{19,
   author = {Lau, Chun Ning and Bockrath, Marc W. and Mak, Kin Fai and Zhang, Fan},
   title = {Reproducibility in the fabrication and physics of moiré materials},
   journal = {Nature},
   volume = {602},
   number = {7895},
   pages = {41-50},
   ISSN = {1476-4687},
   DOI = {10.1038/s41586-021-04173-z},
   url = {https://doi.org/10.1038/s41586-021-04173-z},
   year = {2022},
   type = {Journal Article}
}

@article{20,
   author = {Yu, Hongyi and Liu, Gui-Bin and Tang, Jianju and Xu, Xiaodong and Yao, Wang},
   title = {Moiré excitons: From programmable quantum emitter arrays to spin-orbit–coupled artificial lattices},
   journal = {Science Advances},
   volume = {3},
   number = {11},
   pages = {e1701696},
   DOI = {doi:10.1126/sciadv.1701696},
   url = {https://www.science.org/doi/abs/10.1126/sciadv.1701696},
   year = {2017},
   type = {Journal Article}
}

@article{21,
   author = {Baek, H. and Brotons-Gisbert, M. and Koong, Z. X. and Campbell, A. and Rambach, M. and Watanabe, K. and Taniguchi, T. and Gerardot, B. D.},
   title = {Highly energy-tunable quantum light from moire-trapped excitons},
   journal = {Science Advances},
   volume = {6},
   number = {37},
   pages = {eaba8526},
   DOI = {doi:10.1126/sciadv.aba8526},
   url = {https://www.science.org/doi/abs/10.1126/sciadv.aba8526},
   year = {2020},
   type = {Journal Article}
}

@article{22,
   author = {Ray, Arnab Barman and Mukherjee, Arunabh and Qiu, Liangyu and Sailus, Renee and Tongay, Sefaattin and Vamivakas, Anthony Nickolas},
   title = {Interplay of Trapped Species and Absence of Electron Capture in Moiré Heterobilayers},
   journal = {Nano Letters},
   volume = {23},
   number = {13},
   pages = {5989-5994},
   ISSN = {1530-6984},
   DOI = {10.1021/acs.nanolett.3c01177},
   url = {https://doi.org/10.1021/acs.nanolett.3c01177},
   year = {2023},
   type = {Journal Article}
}

@article{23,
   author = {Datta, Biswajit and Khatoniar, Mandeep and Deshmukh, Prathmesh and Thouin, Félix and Bushati, Rezlind and De Liberato, Simone and Cohen, Stephane Kena and Menon, Vinod M.},
   title = {Highly nonlinear dipolar exciton-polaritons in bilayer MoS2},
   journal = {Nature Communications},
   volume = {13},
   number = {1},
   pages = {6341},
   ISSN = {2041-1723},
   DOI = {10.1038/s41467-022-33940-3},
   url = {https://doi.org/10.1038/s41467-022-33940-3},
   year = {2022},
   type = {Journal Article}
}

@article{24,
   author = {Emmanuele, R. P. A. and Sich, M. and Kyriienko, O. and Shahnazaryan, V. and Withers, F. and Catanzaro, A. and Walker, P. M. and Benimetskiy, F. A. and Skolnick, M. S. and Tartakovskii, A. I. and Shelykh, I. A. and Krizhanovskii, D. N.},
   title = {Highly nonlinear trion-polaritons in a monolayer semiconductor},
   journal = {Nature Communications},
   volume = {11},
   number = {1},
   pages = {3589},
   ISSN = {2041-1723},
   DOI = {10.1038/s41467-020-17340-z},
   url = {https://doi.org/10.1038/s41467-020-17340-z},
   year = {2020},
   type = {Journal Article}
}

@article{25,
   author = {Zhang, Long and Wu, Fengcheng and Hou, Shaocong and Zhang, Zhe and Chou, Yu-Hsun and Watanabe, Kenji and Taniguchi, Takashi and Forrest, Stephen R. and Deng, Hui},
   title = {Van der Waals heterostructure polaritons with moiré-induced nonlinearity},
   journal = {Nature},
   volume = {591},
   number = {7848},
   pages = {61-65},
   ISSN = {1476-4687},
   DOI = {10.1038/s41586-021-03228-5},
   url = {https://doi.org/10.1038/s41586-021-03228-5},
   year = {2021},
   type = {Journal Article}
}

@article{26,
   author = {Guo, Xichen and Ge, Hao and Watanabe, Kenji and Taniguchi, Takashi and Chen, Zhanghai and Gu, Jie},
   title = {Mott Insulator Polariton in a MoSe2/WS2 Moiré Lattice},
   journal = {Nano Letters},
   ISSN = {1530-6984},
   DOI = {10.1021/acs.nanolett.5c04233},
   url = {https://doi.org/10.1021/acs.nanolett.5c04233},
   year = {2025},
   type = {Journal Article}
}

@article{27,
   author = {Villafañe, Viviana and Kremser, Malte and Hübner, Ruven and Petrić, Marko M. and Wilson, Nathan P. and Stier, Andreas V. and Müller, Kai and Florian, Matthias and Steinhoff, Alexander and Finley, Jonathan J.},
   title = {Twist-Dependent Intra- and Interlayer Excitons in Moir\'e ${\mathrm{MoSe}}_{2}$ Homobilayers},
   journal = {Physical Review Letters},
   volume = {130},
   number = {2},
   pages = {026901},
   DOI = {10.1103/PhysRevLett.130.026901},
   url = {https://link.aps.org/doi/10.1103/PhysRevLett.130.026901},
   year = {2023},
   type = {Journal Article}
}

@article{28,
   author = {Glazov, Mikhail M. and Chernikov, Alexey},
   title = {Breakdown of the Static Approximation for Free Carrier Screening of Excitons in Monolayer Semiconductors},
   journal = {physica status solidi (b)},
   volume = {255},
   number = {12},
   pages = {1800216},
   ISSN = {0370-1972},
   DOI = {https://doi.org/10.1002/pssb.201800216},
   url = {https://onlinelibrary.wiley.com/doi/abs/10.1002/pssb.201800216},
   year = {2018},
   type = {Journal Article}
}

@article{29,
  title = {Ultrafast polariton relaxation dynamics in an organic semiconductor microcavity},
  author = {Virgili, T. and Coles, D. and Adawi, A. M. and Clark, C. and Michetti, P. and Rajendran, S. K. and Brida, D. and Polli, D. and Cerullo, G. and Lidzey, D. G.},
  journal = {Phys. Rev. B},
  volume = {83},
  issue = {24},
  pages = {245309},
  numpages = {6},
  year = {2011},
  month = {Jun},
  publisher = {American Physical Society},
  doi = {10.1103/PhysRevB.83.245309},
  url = {https://link.aps.org/doi/10.1103/PhysRevB.83.245309}
}

@article{30,
   author = {Dovzhenko, D. S. and Ryabchuk, S. V. and Rakovich, Yu P. and Nabiev, I. R.},
   title = {Light–matter interaction in the strong coupling regime: configurations, conditions, and applications},
   journal = {Nanoscale},
   volume = {10},
   number = {8},
   pages = {3589-3605},
   ISSN = {2040-3364},
   DOI = {10.1039/C7NR06917K},
   url = {http://dx.doi.org/10.1039/C7NR06917K},
   year = {2018},
   type = {Journal Article}
}

@article{31,
  title = {Theory of intervalley Coulomb interactions in monolayer transition-metal dichalcogenides},
  author = {Dery, Hanan},
  journal = {Phys. Rev. B},
  volume = {94},
  issue = {7},
  pages = {075421},
  numpages = {7},
  year = {2016},
  month = {Aug},
  publisher = {American Physical Society},
  doi = {10.1103/PhysRevB.94.075421},
  url = {https://link.aps.org/doi/10.1103/PhysRevB.94.075421}
}

@article{32,
  title = {Coulomb-engineered heterojunctions and dynamical screening in transition metal dichalcogenide monolayers},
  author = {Steinke, C. and Wehling, T. O. and R\"osner, M.},
  journal = {Phys. Rev. B},
  volume = {102},
  issue = {11},
  pages = {115111},
  numpages = {10},
  year = {2020},
  month = {Sep},
  publisher = {American Physical Society},
  doi = {10.1103/PhysRevB.102.115111},
  url = {https://link.aps.org/doi/10.1103/PhysRevB.102.115111}
}

@article{33,
  title = {Electrostatic screening of free charge-neutral dipoles/excitons in two-dimensional media},
  author = {Xiao, Ke and Kan, Chi-Ming and Fan, Feng-Ren and Parkin, Stuart S. P. and Cui, Xiaodong},
  journal = {Phys. Rev. B},
  volume = {112},
  issue = {16},
  pages = {165429},
  numpages = {10},
  year = {2025},
  month = {Oct},
  publisher = {American Physical Society},
  doi = {10.1103/9fgx-nq97},
  url = {https://link.aps.org/doi/10.1103/9fgx-nq97}
}

@article{34,
   author = {Choi, Junho and Hsu, Wei-Ting and Lu, Li-Syuan and Sun, Liuyang and Cheng, Hui-Yu and Lee, Ming-Hao and Quan, Jiamin and Tran, Kha and Wang, Chun-Yuan and Staab, Matthew and Jones, Kayleigh and Taniguchi, Takashi and Watanabe, Kenji and Chu, Ming-Wen and Gwo, Shangjr and Kim, Suenne and Shih, Chih-Kang and Li, Xiaoqin and Chang, Wen-Hao},
   title = {Moiré potential impedes interlayer exciton diffusion in van der Waals heterostructures},
   journal = {Science Advances},
   volume = {6},
   number = {39},
   pages = {eaba8866},
   DOI = {doi:10.1126/sciadv.aba8866},
   url = {https://www.science.org/doi/abs/10.1126/sciadv.aba8866},
   year = {2020},
   type = {Journal Article}
}

@article{35,
  title = {Nonlinear dynamics of a dense two-dimensional dipolar exciton gas},
  author = {Rapaport, Ronen and Chen, Gang and Simon, Steven H.},
  journal = {Phys. Rev. B},
  volume = {73},
  issue = {3},
  pages = {033319},
  numpages = {4},
  year = {2006},
  month = {Jan},
  publisher = {American Physical Society},
  doi = {10.1103/PhysRevB.73.033319},
  url = {https://link.aps.org/doi/10.1103/PhysRevB.73.033319}
}

@article{36,
   author = {Ivanov, A. L.},
   title = {Quantum diffusion of dipole-oriented indirect excitons in coupled quantum wells},
   journal = {Europhysics Letters},
   volume = {59},
   number = {4},
   pages = {586},
   ISSN = {0295-5075},
   DOI = {10.1209/epl/i2002-00144-3},
   url = {https://dx.doi.org/10.1209/epl/i2002-00144-3},
   year = {2002},
   type = {Journal Article}
}

@article{37,
   author = {Ray, Arnab Barman and Ollis, Trevor and Sethuraj, K. R. and Vamivakas, Anthony Nickolas},
   title = {Diffusion of Valley-Coherent Dark Excitons in a Large-Angle Incommensurate Moiré Homobilayer},
   journal = {Nano Letters},
   volume = {25},
   number = {12},
   pages = {4995-5002},
   ISSN = {1530-6984},
   DOI = {10.1021/acs.nanolett.5c00456},
   url = {https://doi.org/10.1021/acs.nanolett.5c00456},
   year = {2025},
   type = {Journal Article}
}

@article{38,
  title = {Temperature dependence of the cavity-polariton mode splitting in a semiconductor microcavity},
  author = {Pratt, A. R. and Takamori, T. and Kamijoh, T.},
  journal = {Phys. Rev. B},
  volume = {58},
  issue = {15},
  pages = {9656--9658},
  numpages = {0},
  year = {1998},
  month = {Oct},
  publisher = {American Physical Society},
  doi = {10.1103/PhysRevB.58.9656},
  url = {https://link.aps.org/doi/10.1103/PhysRevB.58.9656}
}

@article{39,
   author = {Grant, Richard T. and Michetti, Paolo and Musser, Andrew J. and Gregoire, Pascal and Virgili, Tersilla and Vella, Eleonora and Cavazzini, Marco and Georgiou, Kyriacos and Galeotti, Francesco and Clark, Caspar and Clark, Jenny and Silva, Carlos and Lidzey, David G.},
   title = {Efficient Radiative Pumping of Polaritons in a Strongly Coupled Microcavity by a Fluorescent Molecular Dye},
   journal = {Advanced Optical Materials},
   volume = {4},
   number = {10},
   pages = {1615-1623},
   ISSN = {2195-1071},
   DOI = {https://doi.org/10.1002/adom.201600337},
   url = {https://doi.org/10.1002/adom.201600337},
   year = {2016},
   type = {Journal Article}
}

@article{40,
   author = {Zhang, Feng and Pei, Jiajie and Baev, Alexander and Samoc, Marek and Ge, Yanqi and Prasad, Paras N. and Zhang, Han},
   title = {Photo-dynamics in 2D materials: Processes, tunability and device applications},
   journal = {Physics Reports},
   volume = {993},
   pages = {1-70},
   ISSN = {0370-1573},
   DOI = {https://doi.org/10.1016/j.physrep.2022.09.005},
   url = {https://www.sciencedirect.com/science/article/pii/S0370157322003441},
   year = {2022},
   type = {Journal Article}
}

@article{41,
   author = {Feng, Y-P and Spector, HAROLD N},
   title = {Scattering of screened excitons by free carriers in semiconducting quantum well structures},
   journal = {IEEE journal of quantum electronics},
   volume = {24},
   number = {8},
   pages = {1659-1663},
   ISSN = {0018-9197},
   year = {1988},
   type = {Journal Article}
}

@article{42,
   author = {Edelstein, Warren S. and Spector, Harold N.},
   title = {Binding energy of screened two-dimensional excitons},
   journal = {Surface Science},
   volume = {224},
   number = {1},
   pages = {581-590},
   ISSN = {0039-6028},
   DOI = {https://doi.org/10.1016/0039-6028(89)90934-5},
   url = {https://www.sciencedirect.com/science/article/pii/0039602889909345},
   year = {1989},
   type = {Journal Article}
}

@article{43,
  author  = {Li, Weijie and Lu, Xin and Wu, Jiatian and Srivastava, Ajit},
  title   = {Optical control of the valley Zeeman effect through many-exciton interactions},
  journal = {Nature Nanotechnology},
  volume  = {16},
  number  = {2},
  pages   = {148--152},
  year    = {2021},
  doi     = {10.1038/s41565-020-00804-0},
  url     = {https://doi.org/10.1038/s41565-020-00804-0}
}

@article{44,
  title={Observation of diffusion and drift of the negative trions in monolayer WS2},
  author={Cheng, Guanghui and Li, Baikui and Jin, Zijing and Zhang, Meng and Wang, Jiannong},
  journal={Nano Letters},
  volume={21},
  number={14},
  pages={6314--6320},
  year={2021},
  publisher={ACS Publications}
}

@article{45,
  title={Neutral exciton diffusion in monolayer MoS2},
  author={Uddin, Shiekh Zia and Kim, Hyungjin and Lorenzon, Monica and Yeh, Matthew and Lien, Der-Hsien and Barnard, Edward S and Htoon, Han and Weber-Bargioni, Alexander and Javey, Ali},
  journal={ACS nano},
  volume={14},
  number={10},
  pages={13433--13440},
  year={2020},
  publisher={ACS Publications}
}

@article{46,
  title={Strongly correlated photons on a chip},
  author={Reinhard, Andreas and Volz, Thomas and Winger, Martin and Badolato, Antonio and Hennessy, Kevin J and Hu, Evelyn L and Imamo{\u{g}}lu, Ata{\c{c}}},
  journal={Nature Photonics},
  volume={6},
  number={2},
  pages={93--96},
  year={2012},
  publisher={Nature Publishing Group}
}

@article{47,
  title={Coherent generation of non-classical light on a chip via photon-induced tunnelling and blockade},
  author={Faraon, Andrei and Fushman, Ilya and Englund, Dirk and Stoltz, Nick and Petroff, Pierre and Vu{\v{c}}kovi{\'c}, Jelena},
  journal={Nature Physics},
  volume={4},
  number={11},
  pages={859--863},
  year={2008},
  publisher={Nature Publishing Group UK London}
}

\end{document}